\documentclass[aps,prd,tightenlines,amsmath,amssymb]{revtex4}%
\usepackage{graphicx}% Include figure files
\usepackage{epsfig}
\usepackage{graphics,color}
\usepackage{amssymb}
\usepackage{amsmath}
\usepackage{latexsym}
\usepackage{hyperref}
\usepackage{amsbsy}
\usepackage{bm}
\usepackage{url}

\begin{document}
\title{Bubble dynamics in a strong first-order quark-hadron transition}
\author{Shuying Zhou$^1$}
\author{Song Shu$^2$}
\author{Hong Mao$^1$}
\email {mao@hznu.edu.cn (corresponding author)}
 \affiliation{1. Department of Physics, Hangzhou Normal University, Hangzhou 311121, China}
 \affiliation{2. Department of Physics and Electronic Science, Hubei University, Wuhan 430062, China}

%\date{}

\begin{abstract}
We investigate the dynamics of a strong first-order quark-hadron transition driven by cubic interaction via homogeneous bubble nucleation in the Friedberg-Lee model. The one-loop effective thermodynamics potential of the model and the critical bubble profiles have been calculated at different temperatures and chemical potentials. By taking the temperature and the chemical potential as the variables, the evolutions of the surface tension, the typical radius of the critical bubble and the shift in the coarse-grained free energy in the presence of a nucleation bubble are obtained and the limit on the reliability of the thin-wall approximation is also addressed accordingly. Our results are compared to those obtained for a weak first-order quark-hadron phase transition, especially the spinodal decomposition is relevant.

\end{abstract}

%\pacs{12.39.Fe,12.39.Ba,12.38.Aw,11.10.Wx}

\maketitle

\section{Introduction}
At sufficiently high temperatures and densities, one expects that normal nuclear matter undergoes a phase transition to quark-gluon plasma (QGP), where quarks and gluons become deconfined and essentially chiral. This is a topic of great interest related to the physics of heavy-ion collisions at ultrarelativistic energies as well as to the astrophysics of neutron stars\cite{Yagi:2005yb,Fukushima:2010bq,Braun-Munzinger:2015hba}. Quantum chromodynamics (QCD) as a theory of strong interaction is applicable to determine the properties of strongly interacting matter at high temperatures and densities, however, because of the phenomenon of asymptotic freedom, the nature of the quark-hadron phase transition nevertheless remains an open question, especially when quark chemical potentials are involved in the practical calculations\cite{Fukushima:2010bq}. Therefore, we still lack the capabilities to describe the low-energy nonperturbative phenomena in the framework of QCD theory and have to resort to effective models to study the nontrivial structure of the QCD vacuum, such as the Nambu-Jona-Lasinio (NJL)\cite{Nambu:1961tp,Vogl:1991qt}, the linear sigma model (LSM)\cite{GellMann:1960np}, or their modernized versions, the Polyakov Nambu-Jona-Lasinio model (PNJL)\cite{Costa:2010zw} and the Polyakov Quark Meson Model (PQM)\cite{Schaefer:2007pw}.

The nature of the QCD phase diagram in the temperature and chemical potential plane has been intensively studied in past decades. Most effective models usually predict a smooth crossover transition at low chemical potential and non-zero temperature, while at high density and low temperature, there is a first-order phase transition for QCD phase transitions. At the endpoint of the first-order phase boundary, there should exist a so-called QCD critical endpoint (CEP)\cite{Luo:2017faz}. How to find and identify the CEP in experiment is the main goals of the beam energy scan (BES) program at Relativistic Heavy-Ion Collider (RHIC)\cite{Aggarwal:2010cw} and the Super-Proton Synchrotron (SPS) facilities\cite{Abgrall:2014xwa}. On the theoretical side, a recent study based on chiral effective models shows that a vast part of the QCD phase diagram is crossover if the quark and meson fluctuations are included via the functional renormalization group\cite{Herbst:2013ail}. But the possibility of a first-order phase transition at large baryon chemical potential is not ruled out from both the experimental and the theoretical point of view. In reality, most descriptions of the equation of state (EoS) of neutron stars with a quark core are undertaken in a hybrid equation of state with a hadron phase connected to a quark phase through a first-order phase transition\cite{Glendenning:2000,Ferreira:2020evu,Xia:2020brt}. Moreover, the properties of hybrid stars with a strong first-order phase transition and their relevance to gravitational wave observations will allow one to probe EoS for matter at extreme circumstances\cite{Cao:2018tzm,Paschalidis:2017qmb}. Besides the quark-hadron phase transition, the first-order phase transition shall also play important roles in the evolution history of the early universe, such as its possible roles in electroweak baryogenesis and dark matter\cite{Trodden:1998ym,Huang:2017kzu}. Recently, a strong first-order phase transition is also taken as a potential source of gravitational waves (GW) which could be measured by future detectors\cite{Ellis:2020awk,Wang:2020jrd}. Especially, the approved Laser Interferometer Space Antenna (LISA) project assigns great importance to the direct detection of the electroweak phase transition through the companion GW signals\cite{Caprini:2019egz}.

In a first-order phase transition, the initial metastable (or false) vacuum decays to the stable vacuum through the nucleation of bubbles larger than a critical size, and the nucleation rate of critical bubbles can be calculated from the microphysics using semiclassical methods in Euclidean thermal field theory\cite{Coleman:1977py,Callan:1977pt,Coleman:1988,Linde:1980tt,Linde:1981zj}. Within this framework, an effective thermodynamic potential in the form of a Landau function with the cubic interaction is an important and useful theoretical tool\cite{Yagi:2005yb}. According to the mean-field theory of phase transitions, the free energy density of the system can be expanded in terms of the parameter near the critical point, we can make a general consideration without going into much detail about the underlying dynamics\cite{Goldenfeld:1992qy}. Therefore, at least in mean-field approximation, the thermodynamical potential of the effective models can be parameterized in the form of a Landau expansion around the equilibrium phase with all terms up to quartic term in the region of the first-order phase transition. This scenario has been adopted to describe the dynamical mechanism of bubble nucleation in a strong first-order cosmological electroweak phase transtion\cite{Enqvist:1991xw} and in a weak first-order quark-hadron phase transition\cite{Scavenius:2000bb,Bessa:2008nw}. The benefit of this kind of parameterization is that it simplifies the effective potential to facilitate the solution of the equation of motion of the critical bubble profile with both numerical and analytical methods.

For a weak first-order quark-hadron phase transition, when the temperature is slightly less than the critical temperature $T_c$, the thermodynamic potential exhibits a local minimum aside from the global minimum, as the temperature decreases to some specific value $T_{sp}$, the local minimum gradually disappears and ends at a point of inflection known as spinodal instability. Hence, the effective potential has no potential barrier for $T<T_{sp}$, and  the shift in the coarse-grained free energy due to the appearance of the critical bubble monotonously decreases with the decreasing of the temperature and should eventually become zero at some specific temperature as shown in Refs.\cite{Scavenius:2000bb,Bessa:2008nw}. Since the weak first-order quark-hadron phase transition has been intensively investigated in the framework of the linear sigma model coupled to quarks\cite{Cao:2018tzm,Scavenius:2000bb,Palhares:2010be,Kroff:2014qxa} and the hybrid model by combining EoS obtained within lattice QCD for the quark phase with that of gas of resonances in hadron phase\cite{Bessa:2008nw}. In this work, a strong first-order quark-hadron phase transition induced by an effective potential with a zero-temperature potential barrier is to be considered, and the Friedberg-Lee (FL) model\cite{Friedberg:1976eg} fulfils the requirement.

The FL model was originally developed to describe the static properties of isolated hadrons and their behaviors at low energy. By taking the hadrons as the baglike soliton solutions in vacuum, the model provides us a very intuitive physical explanation of the confinement in QCD theory. Recently, the model has been also extended to finite temperatures and densities to study the deconfinement phase transition in Refs.\cite{Reinhardt:1985nq,Li:1987wb,Gao:1992zd,Mao:2007gm,Shu:2010xj}. it is worth to point out that the FL model and its descendant model with the chiral symmetry\cite{Mao:2013qu} can only predict a first-order phase transition in the phase diagram, of course, this is disagreement with most predictions demonstrated in effective models and lattice QCD data\cite{Yagi:2005yb,Fukushima:2010bq,Braun-Munzinger:2015hba}. The remedy to this problem is to introduce the Polyakov loop in the models, and the results in Ref.\cite{Jin:2015goa} show that the PQM model indeed gives a prediction of a crossover in the low-density region and a weakly first-order phase transition in the high-density region. However, most of these previous studies focus on the thermodynamic effective potential, the properties of isolated hadrons in thermal medium and the phase diagram, while our current study will concentrate on the dynamics of a strong first-order phase transition via bubble nucleation. Nowadays, the strong first-order phase transition gains more and more attentions
both in the astrophysics of neutron stars and cosmological phase transitions in the early universe, especially when GWs are relevant. Although the quantitative results in this work are model-dependent, the general and qualitative results presented in this work can also be applied to study the bubble dynamics of the first-order phase transitions in various fields driven by cubic interaction, especially beyond the limit on the thin-wall approximation.

The paper is organized as follows. In the following section we briefly describe the Friedberg-Lee model and its effective potential  at finite temperatures and densities. In Sec. III, we give detailed description of homogeneous nucleation and the methods used for both numerical and analytic computations of the critical bubble profiles. Our results and discussions are presented in Sec. IV, while in the last section we give the summary.

\section{Model Formulation}
We start with the Lagrangian of the Friedberg-Lee model for a phenomenological scalar field $\sigma$ interacted with the spin-$\frac{1}{2}$ quark fields $\psi$ of the form\cite{Friedberg:1976eg},
\begin{eqnarray}\label{lagrangian}
\mathcal{L}=\overline{\psi}(i\eth -g
\sigma)\psi+\frac{1}{2}\partial_{\mu}\sigma
\partial^{\mu}\sigma -U(\sigma),
\end{eqnarray}
where the potential, which exhibits a typically first-order phase transition, is parameterized in the form of a Landau expansion with all the terms up to quartic term as
\begin{eqnarray}
U(\sigma)=\frac{1}{2!}a\sigma^2+\frac{1}{3!}b \sigma^3+\frac{1}{4!}c \sigma^4.
\end{eqnarray}
The model parameters $a$, $b$ and $c$ are well chosen such that $b^2 > 3ac$ in order to ensure a local minimum of $U(\sigma)$ at $\sigma=0$ and an global minimum at a relative larger value of the $\sigma$ field
\begin{eqnarray}
\sigma_v=\frac{3|b|}{2c}\left[1+\left[1-\frac{8ac}{3b^2}\right]^{\frac
1 2}\right].
\end{eqnarray}
Usually, the global minimum at $\sigma=\sigma_v$ is interpreted as the physical or true vacuum, whereas the local minimum at $\sigma=0$ represents a metastable vacuum where the condensates vanishes and quarks have zero rest mass. The difference in the potential values of the two vacuum states is defined as the quantity $\varepsilon$. For convenience, in the following discussions, we'd like to take $U(0)=0$, therefore, we have
\begin{eqnarray}
-\varepsilon=\frac{a}{2!}\sigma^2_v+\frac{b}{3!}\sigma^3_v+\frac{c}{4!}\sigma^4_v.
\end{eqnarray}
There is a wide range of the model parameters $a$, $b$, $c$ and $g$ adopted in Refs.\cite{Goldflam:1981tg,Li:1987wb,Gao:1992zd} in order to confront the basic properties of nucleon in vacuum. However, for the problem we discuss here, different sets of values will show similar physical results, hereafter we just take one set of parameters $a=17.70 fm^{-2}$,
$b=-1457.4 fm^{-1}$, $c=20000$ and $g=12.16$, which has been widely used in the literatures.

A convenient framework of studying phase transitions is thermal field theory. Within this framework, the finite temperature effective potential is an important and useful theoretical tool. Keeping only contributions to one-loop order, the effective potential of the Friedberg-Lee model can be computed exactly in closed form following the steps presented in Ref.\cite{Dolan:1973qd}
\begin{eqnarray}\label{potential0}
V_{\mathrm{eff}}(\sigma;T,\mu)=U(\sigma)+V_B(\sigma;T)+V_F(\sigma;T,\mu),
\end{eqnarray}
where $V_B(\sigma;T)$ is the finite temperature contribution from boson loop, and $V_F(\sigma;T,\mu)$ is the finite temperature and density contribution from fermion loop\cite{Mao:2007gm,Dolan:1973qd}. These contribute the following terms in the effective potential
\begin{eqnarray}\label{potential1}
V_B(\sigma;T)=T \int \frac{d^3 \vec{p}}{(2\pi)^3}
\mathrm{ln} \left( 1-e^{-E_{\sigma}/T} \right),
\end{eqnarray}
\begin{eqnarray}\label{potential2}
V_F(\sigma;\beta,\mu)=-2N_f N_c T
\int \frac{d^3 \vec{p}}{(2\pi)^3} \left[ \mathrm{ln} \left( 1+e^{-(E_q-\mu)/T} \right)+\mathrm{ln} \left( 1+e^{-(E_q+\mu)/T} \right) \right],
\end{eqnarray}
in which $N_f=2$, $N_c=3$. $E_{\sigma}=\sqrt{\vec{p}^2+m_{\sigma}^2}$ and $E_q=\sqrt{\vec{p}^2+m_q^2}$ are energies for the $\sigma$ mesons and quarks, in which the constituent quark (antiquark) mass $m_q$ is defined as $m_q=g \sigma$, while the effective mass of scalar meson field is set by $m^2_{\sigma}=a+b \sigma+\frac{c}{2} \sigma^2$. To ensure $m_{\sigma}$ to be positive, in this work we prefer to fix it to the vacuum value.

\begin{figure}
\includegraphics[scale=0.36]{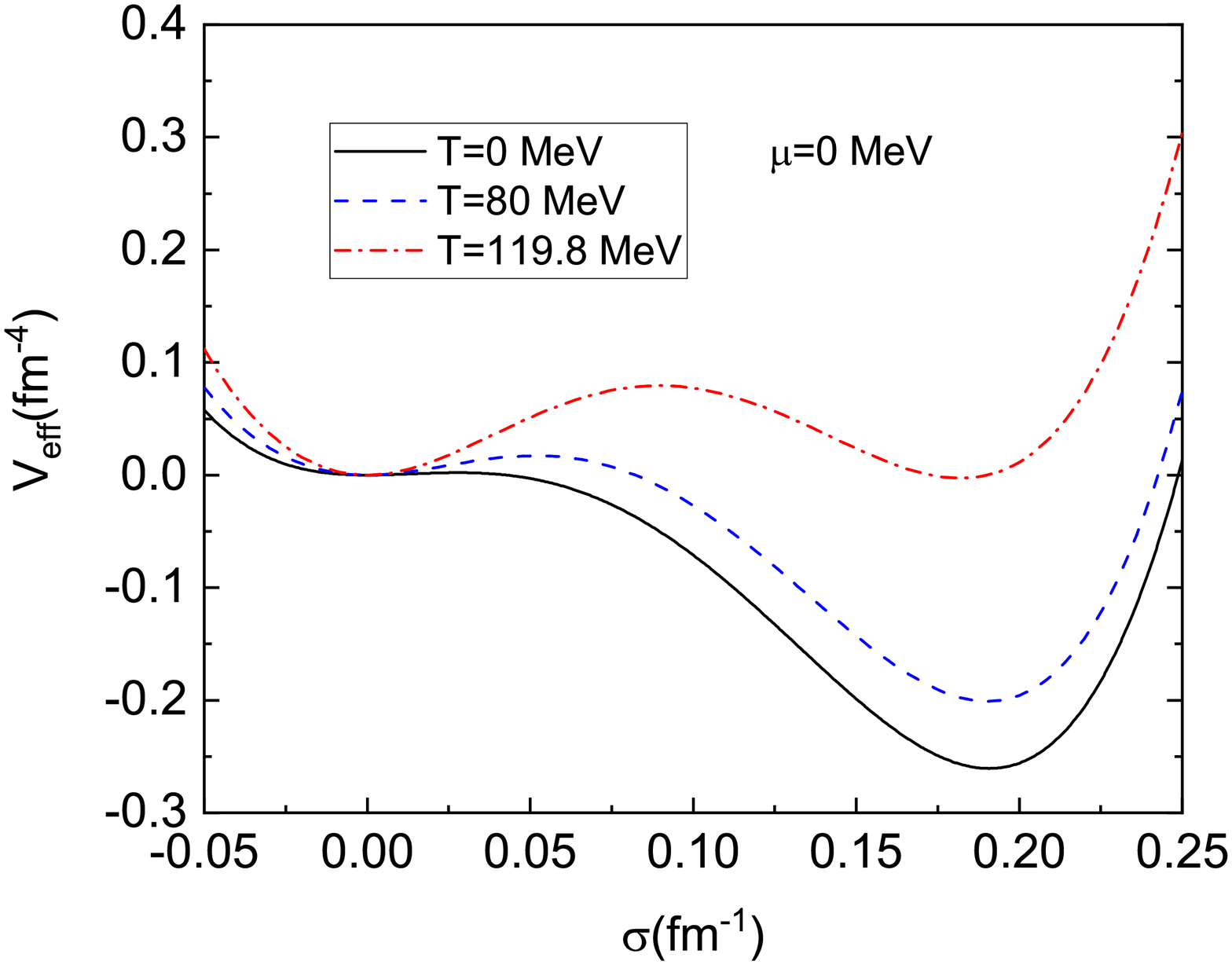}
\caption{\label{Fig:Fig1} The one-loop effective potential $V_{eff}$ as a function of $\sigma$ at $T=0 $ MeV, $T=80$ MeV and $T=119.8$ MeV when fixing the chemical potential $\mu$ at $0$ MeV. For our choice of parameters, the two minima appear as degenerate at $T_{C}\simeq 119.8 \mathrm{MeV}$, which is usually defined as the critical temperature.}
\end{figure}

In the absence of the chemical potential, the one-loop effective potential at different temperatures has been plotted in Fig.\ref{Fig:Fig1}. The shape of the potential shows that a first-order phase transition takes place as it exhibits two degenerate minima at a certain temperature $T_c\simeq 119.8$ MeV, which is usually defined as the critical temperature. Normally, apart from this critical temperature, there exists another particular temperature as one of the minima of the potential disappears when the temperature is at a higher temperature. Between these two particular temperatures, metastable states exist and lie close to $\sigma_v$, and the system can exhibit supercooling or superheating. With temperature decreasing across the critical one, the metastable vacuum and physical vacuum will get flipped, and the metastable states now are centred around the origin $\sigma=0$. Then, the difference between the effective potential at the metastable vacuum state and the physical vacuum state is
\begin{eqnarray}\label{bag1}
\varepsilon(T)=V_{\mathrm{eff}}(0;T)-V_{\mathrm{eff}}(\sigma_v;T).
\end{eqnarray}
It is easy to check that the quantity $\varepsilon$ will decrease with the increasing of the temperature, and when $T=T_c$ the two vacuum are equal, $\varepsilon$ is zero.

\begin{figure}
\includegraphics[scale=0.36]{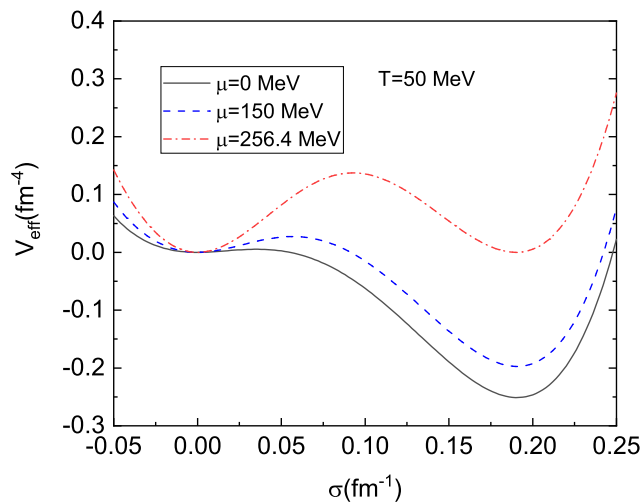}
\caption{\label{Fig:Fig2} The one-loop effective potential $V_{eff}$ as a function of $\sigma$ at $\mu=0$ MeV, $\mu=150$ MeV and $\mu=256.4$ MeV when fixing the temperature $T$ at $50$ MeV. For our choice of parameters, the critical chemical potential is set at $\mu_{C}\simeq 256.4$ MeV when two minima are equal.}
\end{figure}

When fixing temperature at $T=50$ MeV, we plot the one-loop effective potential $V_{\mathrm{eff}}$ as a function of $\sigma$ at various chemical potentials $\mu=0$ MeV, $\mu=150$ MeV and $\mu=256.4$ MeV in Fig.\ref{Fig:Fig2}. From this figure, the shapes of the potentials show similar behaviors as the case in Fig.\ref{Fig:Fig1}. For $\mu=256.4$ MeV, the values of the effective potentials at the two vacuums are equal. In this moment, this certain chemical potential is defined as the critical chemical potential $\mu_c=256.4$ MeV. With the decreasing of the chemical potential from $\mu_c$, the global minimum of the potential jumps from the position of $\sigma=0$ to that of $\sigma_v$. The difference between the values of the effective potential at the false vacuum and at the physical vacuum as usual is defined as
\begin{eqnarray}\label{bag2}
\varepsilon(T,\mu)=V_{\mathrm{eff}}(0;T,\mu)-V_{\mathrm{eff}}(\sigma_v;T,\mu).
\end{eqnarray}
This quantity will also decrease to zero as the increasing of the chemical potential up to its critical value.

\section{Homogeneous thermal nucleation}

For the first-order phase transition, when the temperature or chemical potential is around its critical value, the effective potential exhibits degenerate minima which are separated by a barrier. As the temperature or chemical potential is lowered, the local minimum at $\sigma\simeq 0$ becomes the false vacuum, while the global minimum of the effective potential at $\sigma\approx \sigma_v$ is taken as the stable or physical vacuum. The false vacuum would be stable classically, but quantum mechanically it is only a metastable state and can decay via the nucleation of bubbles larger than a critical size. Technically, this decay may be triggered by either quantum or thermal fluctuations, depending on what kind of physics we are interested in. In this work, we will be mostly concerned with the regime in which thermal fluctuations are much larger than quantum fluctuations.

The dynamics of a first-order phase transition can be described by the mechanism of bubble nucleation of the stable vacuum inside the false vacuum, which is believed to be a natural consequence of the thermal and quantum fluctuations of any thermodynamic systems closely interrelated with a first-order phase transition. For $T<T_c$ or $\mu<\mu_c$, bubbles of the stable vacuums created by thermal fluctuations may grow or shrink inside the false vacuum depending on its energy budget with regard to homogeneous false vacuum. Because the bulk free energy density of the false vacuum is higher than that of the stable vacuum, the phase conversion from the false vacuum to the stable vacuum makes the bulk free energy of the whole system lower. However, the appearance of a spherical bubble means there is an interface which is needed in order to separate the stable vacuum from the exterior of the false vacuum. The creation of such an interface represents an energy cost. Therefore, the mechanism of phase conversion from the metastable phase to the stable phase proceeds by a competition between the free energy gain from the phase transition of the bulk and the energy cost from the formation of an interface. Note that the free energy shift due to the appearance of a spherical bubble of the stable vacuum is proportional to $-R^3$, where $R$ is the bubble radius, and the surface tension of the interface between two phase is proportional to $+R^2$. For the nucleation of small bubbles, the energy cost is higher than the energy gain, small bubbles tends to shrink. On the contrary, a bubble with a sufficiently large radius represents a large bulk energy gain, the energy gain in the system is going to be higher than the surface energy cost in creating the bubble. As a consequence, these large bubbles tend to expand even more and to coalesce completely, completing the phase conversion. Therefore, only bubbles of a very large radius play a decisive role in the theory of dynamics of a first-order phase transition.

In the theory of the bubble nucleation, a scalar field $\sigma$ is treated as the order parameter and a coarse-grained free energy functional of the system is defined as
\begin{equation}\label{fenergy}
F(\sigma)=\int dr^3 \left[ \frac{1}{2} \left(\nabla \sigma \right)^2+V_{\mathrm{eff}}(\sigma;T,\mu)    \right].
\end{equation}
The critical bubble configuration is an extremum of the coarse-grained free energy functional $F(\sigma)$ with respect to the scalar field $\sigma$, so that the equation of motion to be solved is now the nonlinear ordinary differential equation,
\begin{equation}\label{eom}
\frac{d^2\sigma(r)}{dr^2}+\frac{2}{r}\frac{d\sigma(r)}{dr}=\frac{\partial V_{\mathrm{eff}}(\sigma;T,\mu)}{\partial \sigma},
\end{equation}
with boundary conditions $\lim\limits_{r \to \infty }\sigma(r)=0$ and $\frac{d\sigma(0)}{dr}=0$. The first boundary condition is because that the bubbles are embedded in the homogeneous false vacuum, outside the bubble, the $\sigma$ field should arrive at its false vacuum at $\sigma \simeq 0$. While the second one is set by the requirement of no singularity of the solution at the origin. The solution for this equation of motion with the above proper boundary conditions is a saddle point solution $\sigma_b$.

Once the solution $\sigma_b$ is found, the shift in the coarse-grained free energy due to the formation of a nucleation bubble can be calculated as
\begin{equation}\label{fenergyb}
\Delta F_b=4 \pi \int r^2 dr \left[ \frac{1}{2} \left( \frac{d\sigma_b}{dr} \right)^2+V_{\mathrm{eff}}(\sigma_b;T,\mu)  \right].
\end{equation}
It should be pointed out that here and after that $V_{\mathrm{eff}}(0;T,\mu)$ is well normalized to be zero for simplicity. The nucleation rate per unit volume is expressed as\cite{Linde:1980tt,Linde:1981zj}
\begin{equation}\label{rate}
\Gamma=\mathcal{P}\exp\left[ -\frac{\Delta F_b}{T}  \right],
\end{equation}
where the pre-exponential factor $\mathcal{P}$ corresponds to the probability for a critical bubble-like field fluctuation $\sigma_b$ to be generated and grow. Evaluation of the pre-exponential factor is a nontrivial matter. A rough estimate of their ratio can be obtained by dimensional arguments and we could approximate $\mathcal{P}$ by $T^4$ for simplicity\cite{Scavenius:2000bb}. The surface tension of the nucleation bubble interface between the false vacuum and the stable vacuum is then defined as
\begin{equation}\label{surfacet}
\Sigma=\int dr \left[ \frac{1}{2} \left(\frac{d\sigma_b}{dr}  \right)^2+V_{\mathrm{eff}}(\sigma_b;T,\mu)    \right].
\end{equation}

For a generic effective potential $V_{\mathrm{eff}}$, the equation of motion (\ref{eom}) with some certain boundary conditions usually cannot be solved analytically. However, when the system is very close to the critical coexistence line, e.g. $T\sim T_c$ or $\mu \sim \mu_c$, the problem can be essentially simplified. In such a situation, the quantity $\varepsilon$ is much smaller than the height of the barrier separated these two vacua, because of the competition between the free energy gain and the surface energy cost, the typical radius of the bubbles becomes much greater than the wall thickness, the second term in the equation of motion (\ref{eom}) can be neglected. Then the so-called thin-wall approximation is applicable and the equation of motion (\ref{eom}) reduces to the equation for a typical one-dimensional soliton
\begin{equation}\label{eom2}
\frac{d^2 \sigma (r)}{dr^2}=\frac{d V_{\mathrm{eff}}}{d \sigma}.
\end{equation}
This static field equation implies that
\begin{equation}\label{eom3}
\frac{d \sigma(r)}{dr}=\pm \sqrt{2 V_{\mathrm{eff}}}.
\end{equation}
Integrating Eq.(\ref{eom3}) yields
\begin{eqnarray}\label{radius}
  r &=& \int_{\sigma}^{\sigma_v} \frac{d \sigma}{\sqrt{2 V_{\mathrm{eff}}}}.
\end{eqnarray}

In the case of an arbitrary potential $V_{\mathrm{eff}}$ with two or more degenerate global minima as in the limit $\varepsilon\rightarrow 0$, the profile of the critical bubble can be estimated as follows. For a smoothly varying potential $V_{\mathrm{eff}}$, the integral on the right-hand side diverges as $\sigma(r)$ approaches any of the global minima. Hence, as $r$ ranges from $0$ to $\infty$, $\sigma(r)$ must vary monotonically from one global minimum of $V_{\mathrm{eff}}$ at $\sigma=\sigma_v$ to an adjacent global minimum at $\sigma=0$. In this case, the approximate solution for the bubble with the critical size is then given by
\begin{equation}\label{sgmtw}
 \sigma(r)=\begin{cases}
 \sigma_v & 0<r<R-\Delta R, \\
 \sigma_{\mathrm{wall}}(r) & R-\Delta R<r<R+\Delta R, \\
 0 & r>R+\Delta R,
 \end{cases}
 \end{equation}
which indicates that the stable vacuum inside the bubble is separated from the metastable one outside by the bubble wall $\sigma_{\mathrm{wall}}(r)$, solved from Eq.(\ref{radius}). Moreover, in the thin-wall approximation, since there exists an energy competition between the free energy gain and the surface energy cost, the free energy $F(R)$, relative to the false vacuum background, of a bubble with a radius $R$ could be expressed as\cite{Linde:1981zj,Weinberg:2012pjx}
\begin{eqnarray}\label{fenergb2}
F(R)= 4 \pi R^2 \Sigma-\frac{4}{3}\pi R^3 \varepsilon.
\end{eqnarray}
Here, the first term is the contribution from the bubble wall with a surface tension $\Sigma$, while the second is from the true vacuum interior. The typical radius $R_c$ of the bubble is determined by minimization of the free energy $F(R)$ with respect to $R$, which in turn requires that
\begin{equation}
0=\frac{dF}{dR}=8 \pi R \Sigma- 4 \pi R^2 \varepsilon.
\end{equation}
This is solved by
\begin{equation}\label{cradius}
R_c=\frac{2 \Sigma}{\varepsilon}.
\end{equation}
As described in previous discussion, only bubbles that have a size equal to or larger than the typical radius $R_c$ are energetically favourable and would play an important role in the dynamical seed of the phase conversion.

In the last of this section, it is worth to note that, in the absence of the quantity $\varepsilon$, the one-dimensional energy or the surface tension of the bubble is
\begin{equation}\label{surfacet2}
\Sigma_{tw}=\int_{0}^{\infty} dr \left[ \frac{1}{2} \left(\frac{d\sigma_b}{dr} \right)^2+V_{\mathrm{eff}} \right]=\int_{0}^{\sigma_v} d\sigma \sqrt{2 V_{\mathrm{eff}}} .
\end{equation}
From the equations (\ref{radius}) and (\ref{surfacet2}), a saddle point field configuration $\sigma(r)$ and the surface tension can be directly obtained by using the effective potential $V_{\mathrm{eff}}$ without solving the equation of motion in Eq.(\ref{eom}), which is usually difficult to be solved analytically or even numerically. This is the main advantage of the thin-wall approximation approach. Since the thin-wall approximation is so widely adopted in literatures\cite{Scavenius:2000bb,Bessa:2008nw,Palhares:2010be,Kroff:2014qxa,Gleiser:1993hf,Mintz:2012mz,Fraga:2018cvr,Mao:2019aps}, in what follows, we focus our study on the exact numerical computations and establish limits on the reliability of the thin-wall approximation.

\section{Results and discussion}

\begin{figure}[!thb]
	\includegraphics[scale=0.3]{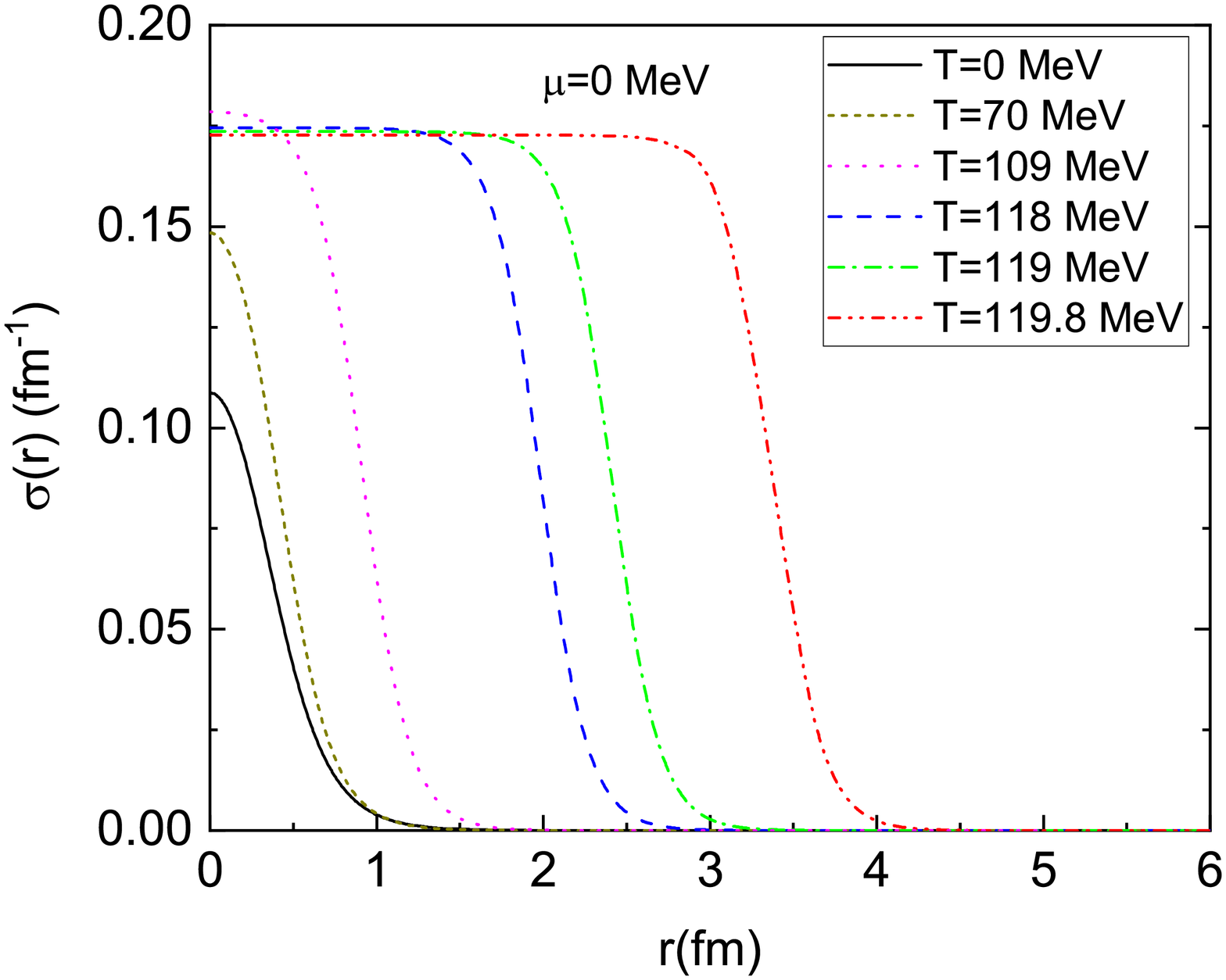}
	\includegraphics[scale=0.3]{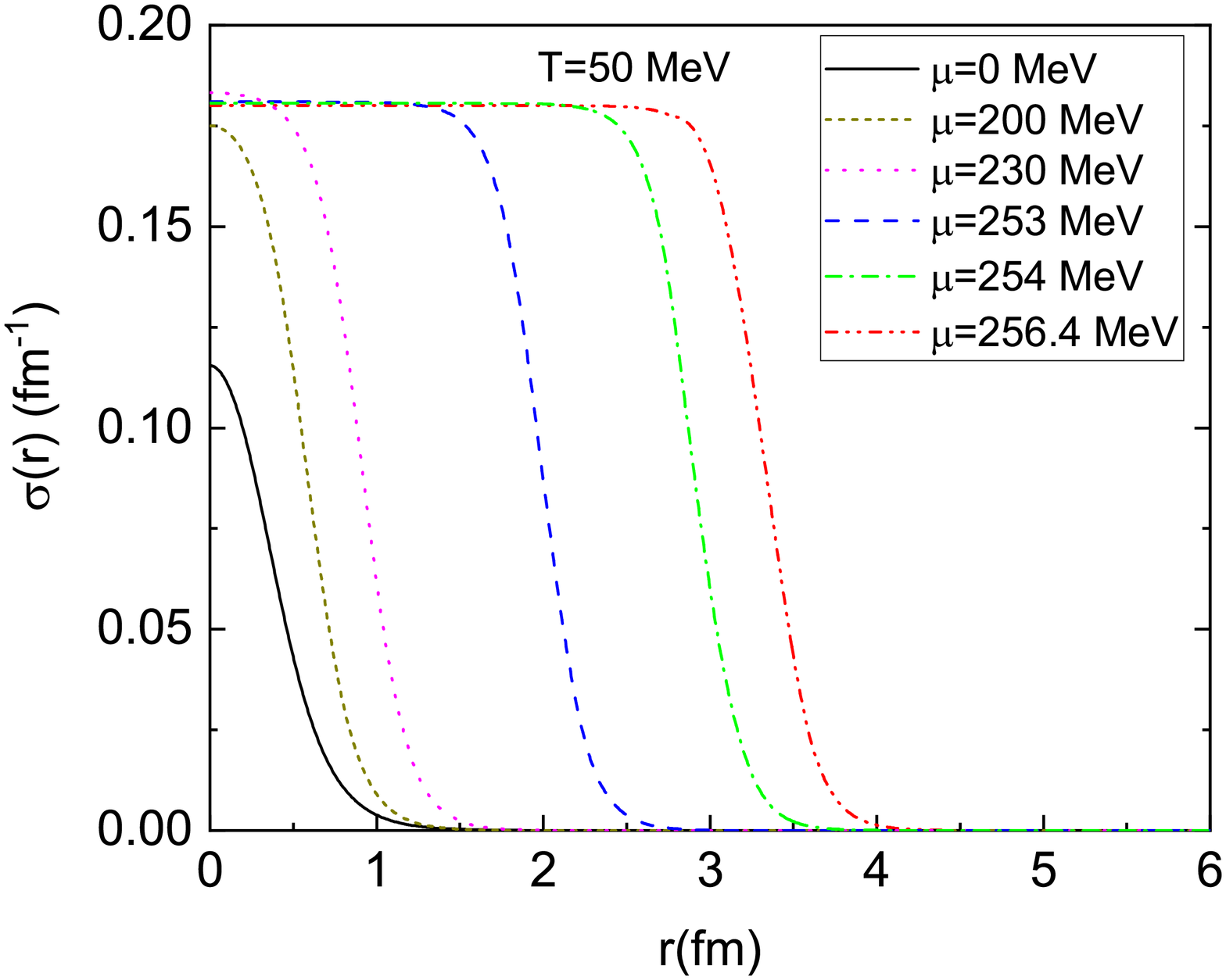}	
	\caption{Left panel: Critical bubble profiles for different temperatures and zero chemical potential. From left to right, the curves correspond to $T=0$, $70$, $109$, $118$, $119$, $119.8$ MeV.
	Right panel: Critical bubble profiles for different chemical potentials when fixing the temperature $T$ at $50$ MeV. From left to right, the curves correspond to $\mu=0$, $200$, $230$, $253$, $254$, $256.4$ MeV.
		\label{Fig:Fig3} }
\end{figure}

In what follows, we numerically solve the equation of motion in Eq.(\ref{eom}) with some proper boundary conditions, $\sigma\rightarrow 0$ as $r\rightarrow \infty$ and $\frac{d \sigma(0)}{dr}=0$. The exact numerical solutions by taking the temperatures as $T=0$, $70$, $109$, $118$, $119$, $119.8$ MeV in the absence of chemical potential are plotted in the left panel of Fig.\ref{Fig:Fig3}. One can see that with the temperature decreasing from $T=T_c$, all curves approach to zero when the radius $r$ is large, whereas $\sigma(r)$ at the center of the bubble is changed dramatically. For the temperature is sufficiently close to the critical temperature at $T_c=119.8$ MeV, the $\sigma$ field at the center of the bubble only slightly deviates from its stable vacuum value at $\sigma=\sigma_v$, however, for $T \leq 109$ MeV the $\sigma$ field at the center of the bubble is visibly different from its stable vacuum value. Such a deviation can be demonstrated by an ``overshoot-undershoot" argument due to Coleman \cite{Coleman:1977py}. According to this idea, the equation of motion (\ref{eom}) is reinterpreted as the equation for a particle moving in an ``upside-down" potential energy $-V_{\mathrm{eff}}$, the $\sigma'(r)$ term is interpreted as a damping force. The boundary conditions require that the particle starts at rest at some initial point $\sigma_0$ on the true vacuum side of the potential well, and it rolls down to rest at its false vacuum $\sigma(0)$. For $-V_{\mathrm{eff}}(\sigma_0)\leq -V_{\mathrm{eff}}(0)$, because of the damping term, the particle will never have sufficient energy to reach $\sigma_0$, it undershoots. On the contrary, if $\sigma_0$ is taken to differ only infinitesimally from $\sigma_v$, the particle could have nonzero kinetic energy when it reaches $\sigma_0$, it will continue on and never return, it overshoots. The desired $\sigma_0$ that determines the bounce is located among these two ranges. In our case, when the temperature is very close to the critical temperature $T_c$, the damping force will have almost died away and the potential has two degenerate vacua, the field $\sigma_0$ starts at the top of the effective potential $-V_{\mathrm{eff}}$ around $\sigma\simeq\sigma_v$. However, with the temperature goes down, two degenerate vacua get decoupled and the damping force takes effect, the field $\sigma_0$ will deviate from its vacuum value more and more dramatically. In other words, the thin-wall approximation is expected to be invalid, and any further extension of the thin-wall approximation to lower temperatures deviation from $T_c$ should be checked very carefully.

Similar discussion can be applied to the second case, when the temperature is fixed, the critical bubble profiles at different chemical potentials are illustrated in the right panel of Fig.\ref{Fig:Fig3}, where the chemical potentials are taken as $\mu=0$, $200$, $230$, $253$, $254$ and $256.4$ MeV for a fixed temperature $T=50$ MeV. The evolution of the $\sigma(r)$ for different chemical potentials tells that the typical radius of the critical bubble should increase as well with the chemical potential increasing, and the nontrivial behavior of the $\sigma(r)$ in the center of the bubble can also be interpreted as a limit to the applicability of the thin-wall approximation. From the right panel of Fig.\ref{Fig:Fig3}, since the $\sigma(0)$ reaches its maximum when $\mu\simeq 230$ MeV, this specific value is taken as the lower limit for the validity of the thin-wall approximation.

\begin{figure}[!thb]
	\includegraphics[scale=0.3]{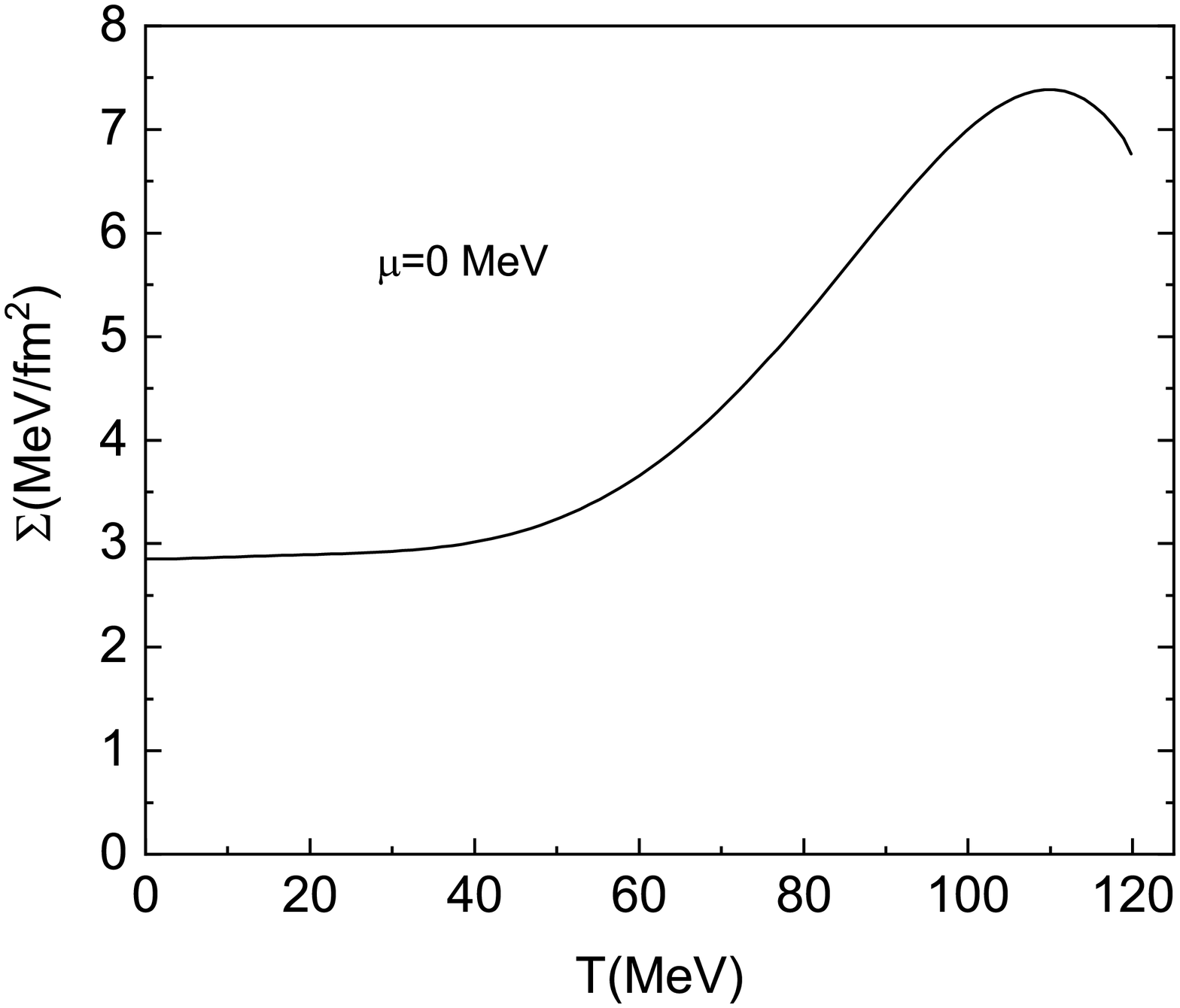}
	\includegraphics[scale=0.3]{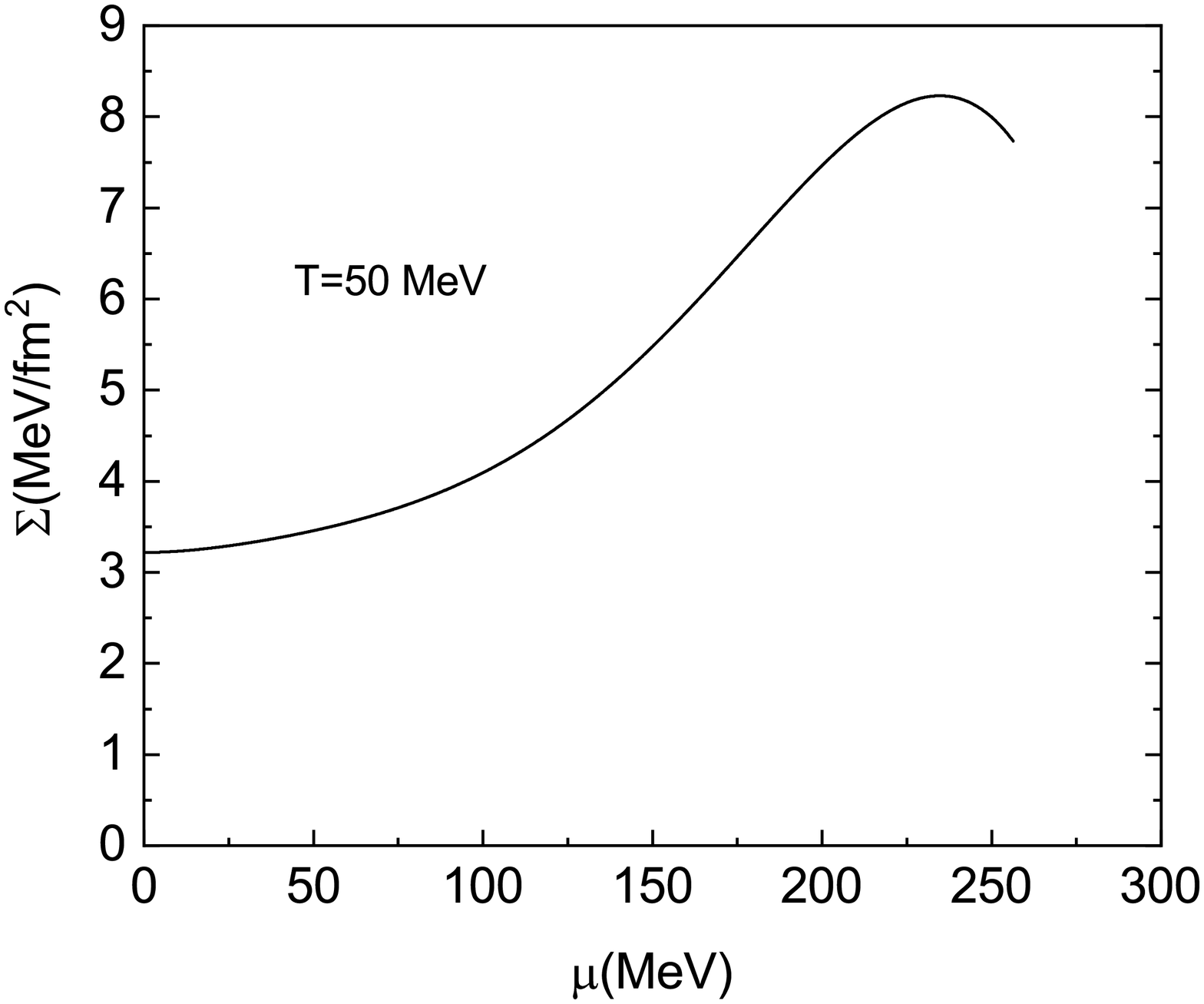}	
	\caption{Left panel: Surface tension as a function of temperature $T$ for $T\leq T_c$ at zero chemical potential.
	Right panel: Surface tension as a function of chemical potential $\mu$ for $\mu \leq \mu_c$ when fixing the temperature at $50$ MeV.
		\label{Fig:Fig4} }
\end{figure}

Once the bubble profiles have been solved, the surface tension of the nucleation bubble interface between the false vacuum and the stable vacuum as a function of the temperature is shown in the left panel of Fig.\ref{Fig:Fig4} in the case of zero chemical potential. An interesting behavior is found: with the increasing of the temperature, the surface tension $\Sigma(T)$ starts to grow quickly from $T=60$ MeV and reaches a maximum $\Sigma(T)\simeq 7.38$ $\mathrm{MeV}/\mathrm{fm}^2$ at $T\simeq 109$ MeV. This nontrivial behavior of $\Sigma(T)$ at $T\simeq 109$ MeV can be analysed by the evolution of the bubble profile with the temperature. From the left panel of Fig.\ref{Fig:Fig3}, as $T$ goes down from its critical temperature $T_c$, the $\sigma(r)$ field nearby the center of the bubble will departure from its thin-wall approximate solution $\sigma_v$ in Eq.(\ref{sgmtw}) gradually, when $T\simeq 109$ MeV, the $\sigma(r)$ field reaches its maximal value before it starts to decrease. This implies that the turning point of the surface tension could be treated as a landmark for the breaking down of the thin-wall approximation. For the second case presented in the right panel of Fig.\ref{Fig:Fig4}, $\Sigma(\mu)$ shows a similar behavior. With the increase of the chemical potential, $\Sigma(\mu)$ goes up accordingly until it reaches the top of its values as $\mu\simeq 230$ MeV, then it drops quickly to some small values. The turning point of is also treated as a generous limit to the applicability of the thin-wall approximation. It is worth to note that the non-monotonic behavior of the surface tension in the present work is also reported in the case of a weak first-order phase transition \cite{Bessa:2008nw}, where the evolution of the surface tension firstly increases to its maximum value, then it decreases rapidly to zero, rather than to a small value. This is the main difference between the strong first-order phase transition and the weak ones. The reason is that for a weak first-order phase transition, as long as the temperature is under a spinodal temperature $T_{sp}$, a small barrier between the two minima in the potential will disappear, and there is only one minimum left in the effective potential. According to a standard criterion to guarantee the existence of the stable bounce, it is indispensable for the potential of the order parameter fields, e.g. $\sigma$ field in this work, to exhibit three distinct extrema \cite{Coleman:1988, Weinberg:2012pjx,Goldflam:1981tg,Jin:2015goa}. So that we can only have a trivial solution to the equation of motion (\ref{eom}) as $\sigma(r)=0$ if $T<T_{sp}$, and the surface tension should approach to the zero when $T\rightarrow T_{sp}$.

\begin{figure}[!thb]
	\includegraphics[scale=0.3]{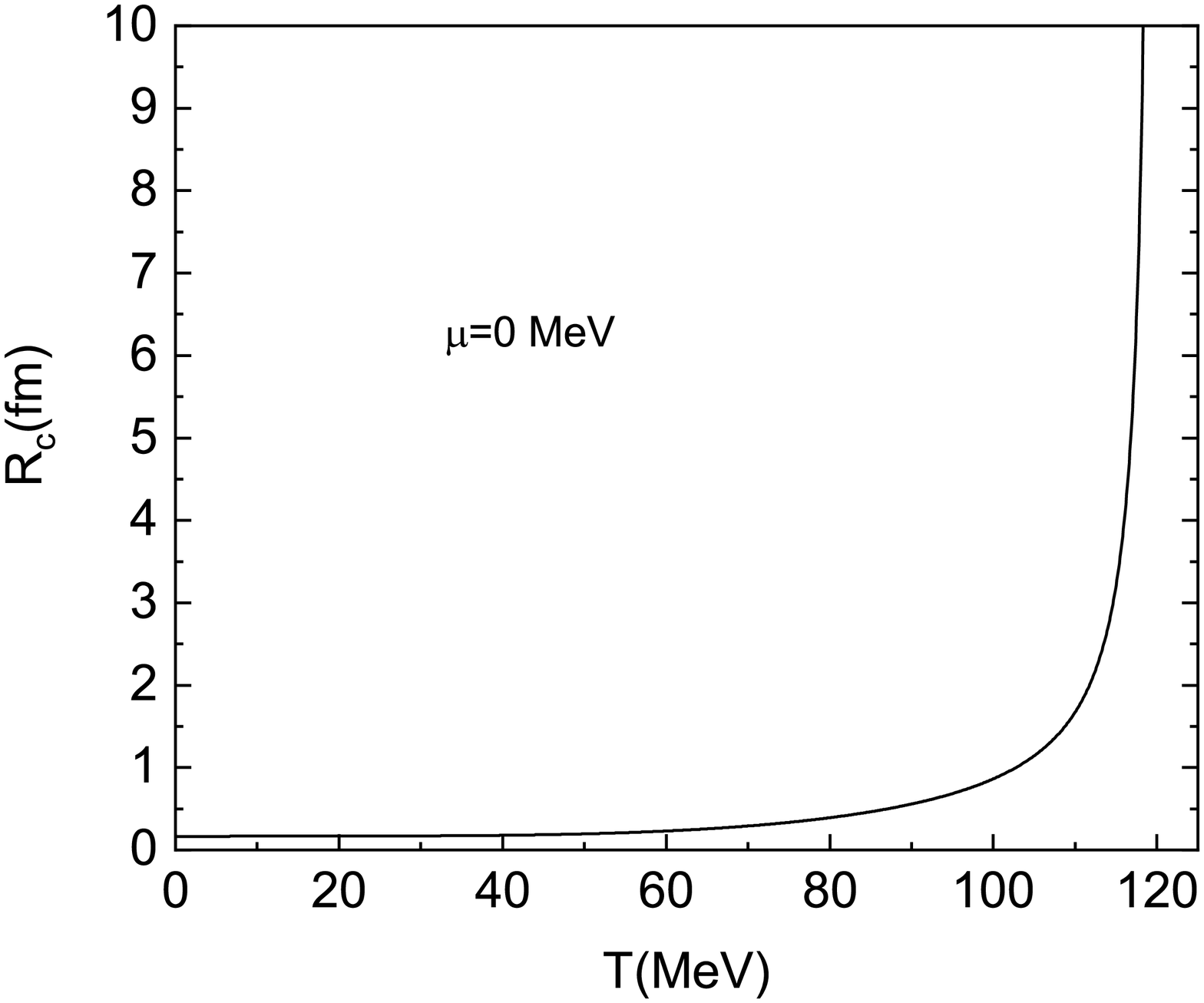}
	\includegraphics[scale=0.3]{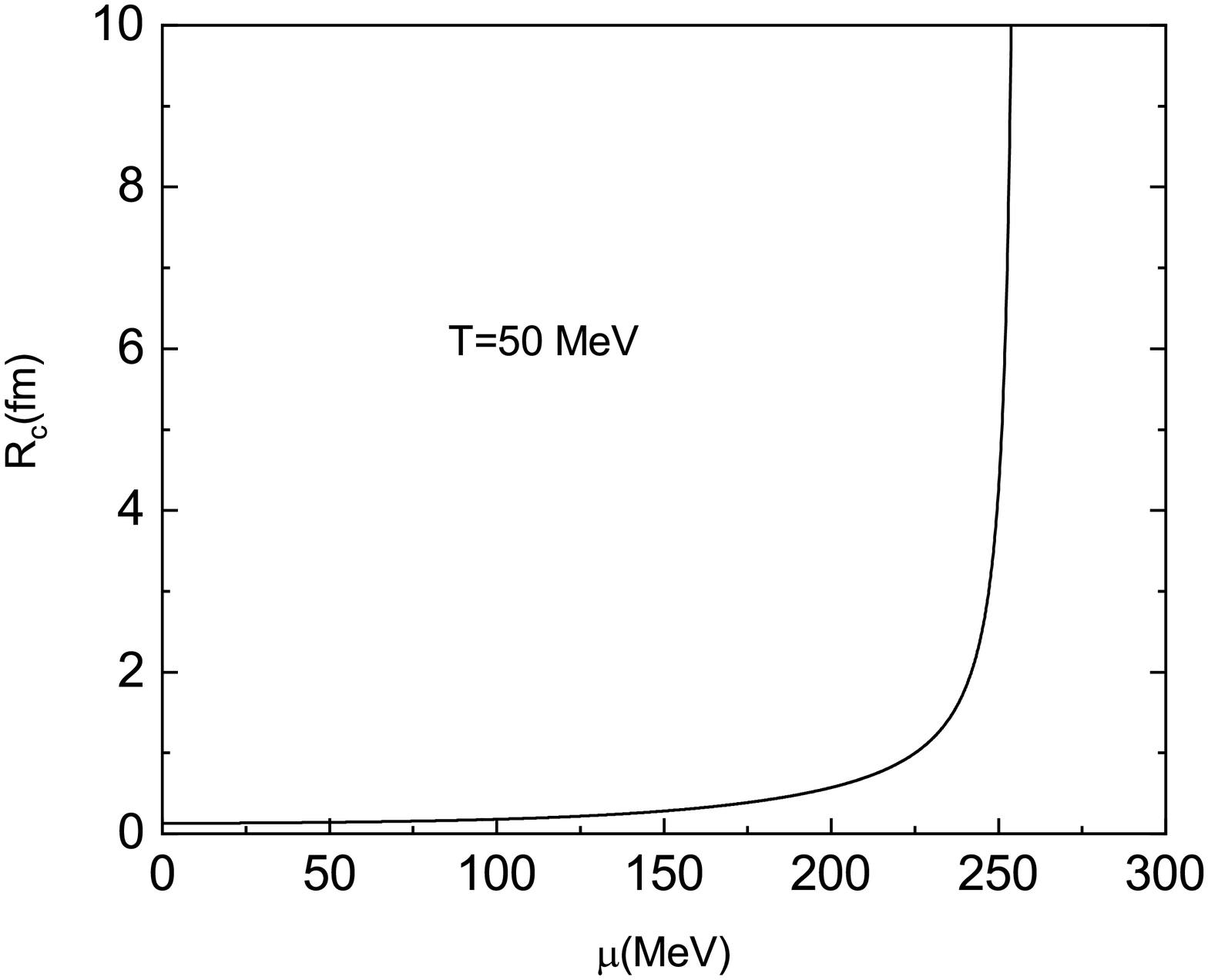}	
	\caption{Left panel: The typical radius of the critical bubble as a function of temperature $T$ when $T\leq T_c$ at zero chemical potential.
	Right panel: The typical radius of the critical bubble as a function of chemical potential $\mu$ when $\mu \leq \mu_c$ for fixing the temperature at $50$ MeV.
		\label{Fig:Fig5} }
\end{figure}

The typical radius of the critical bubble as a function of temperature and chemical potential are displayed in Fig.\ref{Fig:Fig5}. As mentioned above, any bubble smaller than the critical bubble will shrink and rapidly disappear, and any larger bubble will grow and drive the phase conversion. Therefore, bubbles with radii larger than $R_c$ will have a decisive role and can be taken as the dynamical seed of the first-order phase conversion.

From Fig.\ref{Fig:Fig5}, the critical bubble swells with the increase of temperature and chemical potential, more obvious for larger variables, and diverges at $T=T_c$ and $\mu=\mu_c$. The divergent behaviors of the $R_c$ at $T=T_c$ and $\mu=\mu_c$ are in agreement with the definition of the typical radius in Eq.(\ref{cradius}), note that $\varepsilon\rightarrow 0$ as $T \rightarrow T_c$. For the numerical subtlety, besides the nontrivial numerical solutions presented in Fig.\ref{Fig:Fig3}, the equation of motion in Eq.(\ref{eom}) can always possess two trivial solutions: $\sigma(r)=\sigma_v$ and $\sigma(r)=0$. The former trivial solution is subject to the divergence of the $R_c$ when the system is at its critical point.

\begin{figure}[!thb]
	\includegraphics[scale=0.3]{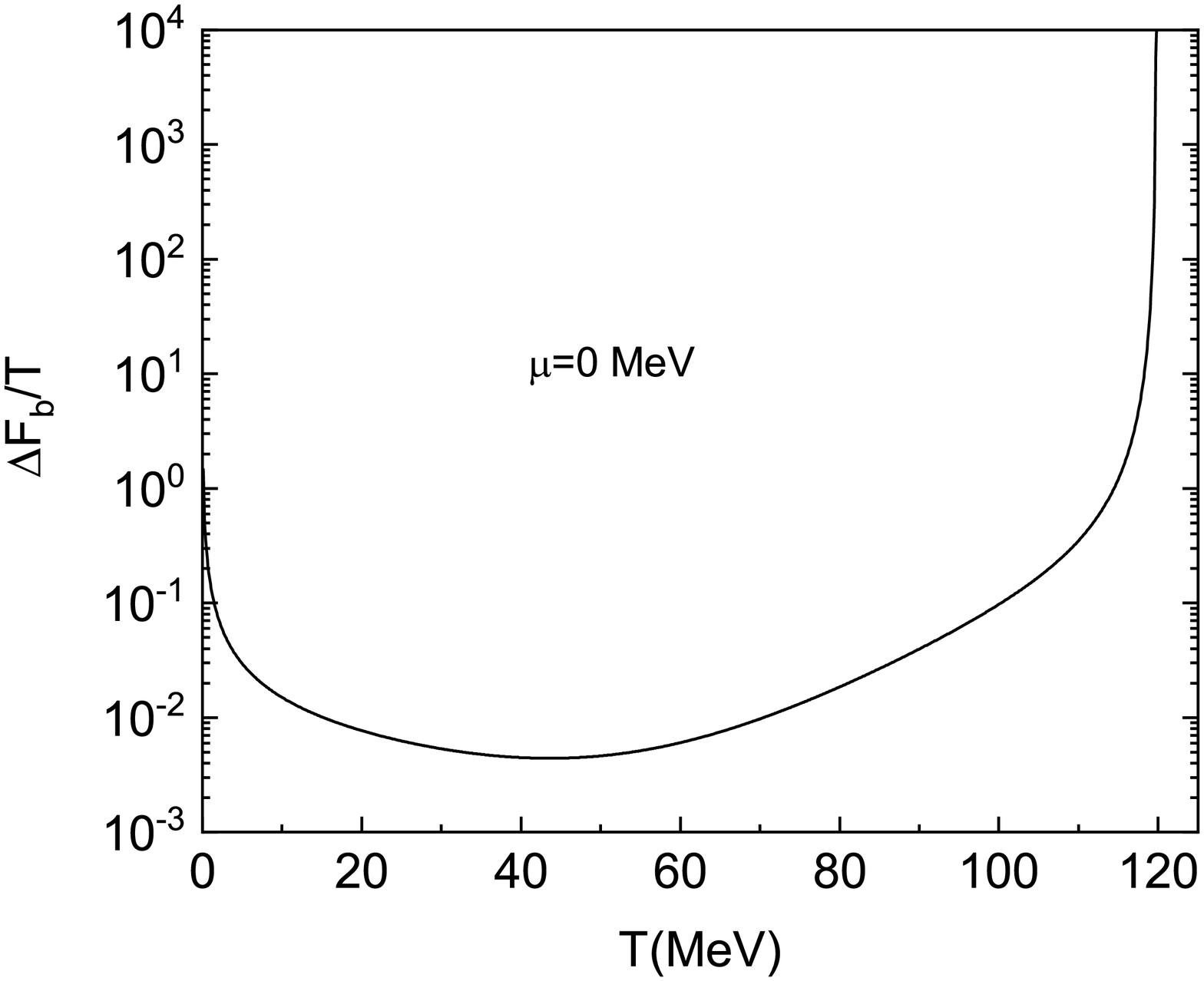}
	\includegraphics[scale=0.3]{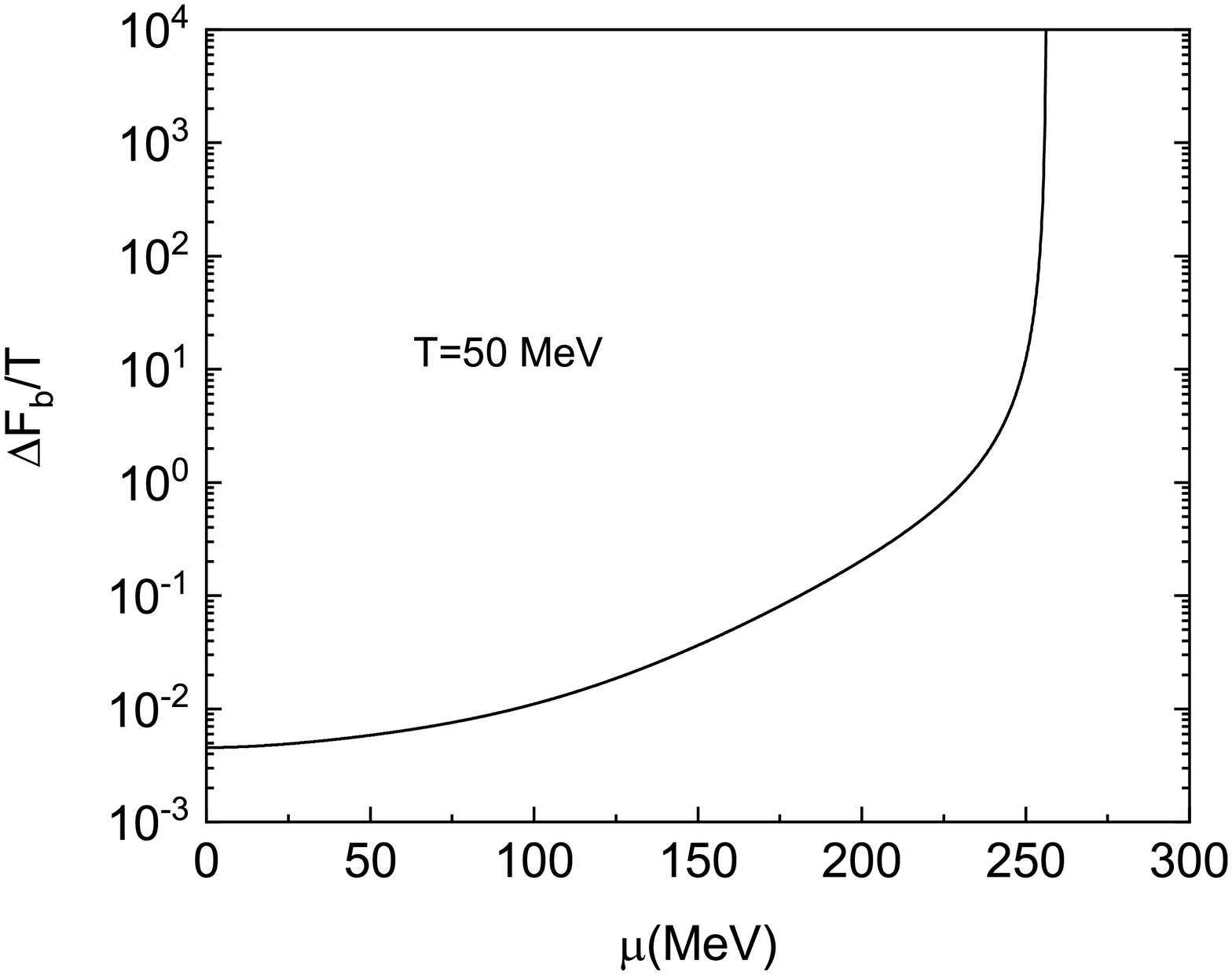}	
	\caption{Left panel: The bubble activation free energy shift $\Delta F_b/T$ as a function of temperature $T$ for $T\leq T_c$ at zero chemical potential.
	Right panel: The bubble activation free energy shift $\Delta F_b/T$ as a function of chemical potential $\mu$ for $\mu \leq \mu_c$ when fixing the temperature at $50$ MeV.
		\label{Fig:Fig6} }
\end{figure}

The shift in the coarse-grained free energy due to the activation of a nucleation bubble $\Delta F_b$ can be calculated directly from Eq.(\ref{fenergyb}). In this work, we concentrate on the relatively violent behavior of the exponential factor in Eq.(\ref{rate}), which is an essential ingredient for the nucleation rate per unit volume $\Gamma$, whereas the pre-exponential factor $\mathcal{P}$ is crudely chosen as $T^4$. To show the shift in the coarse-grained free energy due to the appearance of the critical bubble and its crucial role played in the nucleation rate for the first-order phase transition, $\Delta F_b/T$ as functions of the temperature $T$ and the chemical potential $\mu$ are plotted in Fig.\ref{Fig:Fig6}. In the absence of the chemical potential, the $\Delta F_b/T$ decreases with the increase of the temperature and touches down some minimum point, then it will rise very quickly and diverge nearby the critical temperature $T_c$. For $T\simeq114.5$ MeV, $\Delta F_b/T \simeq1$, then $\Gamma$ will be strongly suppressed by the exponential factor, and the system is likely to stay in the metastable vacuum for a relatively long time. On the contrary, for $T<114.5$ MeV, the unstable vacuum tends to decay very quickly to the true vacuum. The non-monotonic behavior of the $\Delta F_b/T$ as functions of the temperature $T$ is also reported in a recent work on the bounce action for a strong cosmological first-order phase transition \cite{Wang:2020jrd}. As fixing the temperature at $50$ MeV, $\Delta F_b/T$ as a function of chemical potential $\mu$ for $\mu \leq \mu_c$ is also addressed in the right panel in Fig.\ref{Fig:Fig6}. In this case, when $\mu$ is about $231$ MeV, $\Delta F_b/T\simeq1$, so that the system is likely to remain in the metastable vacuum as long as the chemical potential is larger than $231$ MeV.

\section{Summary}
In the present paper we have investigated a dynamics of a strong first-order phase transition via homogeneous bubble nucleation within the Friedberg-Lee model at finite temperatures and chemical potentials. After obtaining the effective thermodynamical potential, a saddle point solution of the equation of motion and the exact bubble profiles have been numerically calculated. For zero chemical potential, the critical temperature $T_c$ is around $119.8$ MeV when the two minima of the effective potential are equal to each other. Alternatively, when taking the chemical potential as a variable, a critical chemical potential is to be set up at $\mu\simeq 256.4$ MeV as temperature is fixed at $50$ MeV.

The evolution of surface tensions in thermal medium show similar behaviors. They will firstly increase to a maximum value and then decrease with the decreases of the temperature or chemical potential. The top of the surface tension can be taken as a limit on the reliability of the thin-wall approximation, since the bubble profile in this point represents a large distortion of that of the thin-wall approximation. Moreover, since two minima of the classical potential in the Friedberg-Lee model are separated by a barrier, no matter how small the barrier will be, we can always have a nontrivial bounce solution for the equation of motion of the bubble profiles. This indicates that as the temperature or the chemical potential goes to zero, the surface tension $\Sigma$ shall approach to a small value rather than zero as long as the barrier exists there. On the contrary, for a weak first-order phase transition, since the local minimum of the effective potential will gradually disappear when $T=T_{sp}$, there only exits a trivial bounce solution for the equation of motion, the surface tension $\Sigma$ should subsequently become zero in this moment\cite{Bessa:2008nw}. This is an apparent feature between the zero-temperature effective potential with and without a barrier.
Furthermore, because of its important role in heavy-ion collision and in astrophysics, the surface tension has attracted much attention recently. Most effective models predict $\Sigma\leq 30$ $\mathrm{MeV/fm^2}$, such as the MIT bag model\cite{Oertel:2008wr}, the quark-meson model\cite{Palhares:2010be,Kroff:2014qxa,Mao:2019aps,Pinto:2012aq}, NJL model\cite{Garcia:2013eaa,Ke:2013wga}, three-flavor PQM model\cite{Mintz:2012mz}, the nucleon-meson model\cite{Fraga:2018cvr}. Our calculations give rather low values, for example, it is about $7.38$ $\mathrm{MeV/fm^2}$ for zero chemical potential and $7.73$ $\mathrm{MeV/fm^2}$ for fixing the temperature at $T=50$ MeV when the system is at the critical point.

Unlike the surface tension, the typical radius of the critical bubble exhibits a monotonic property with the increase of temperature or chemical potential. In both cases, $R_c$ starts from a small value and then increases slightly with the increase of the variable, when the system is close to its critical point, it sharply grows and disappears. However, for a weak first-order quark-hadron phase transition, $R_c\rightarrow 0$ as $T\rightarrow T_{sp}$ because there only has a trivial bounce solution as $T\leq T_{sp}$.

The shift in the coarse-grained free energy $\Delta F_b/T$ show a very interesting behavior with the system warming up. When the temperature rises up, $\Delta F_b/T$ firstly decreases to a minimum value and then increases rapidly. As the temperature is close to the critical temperature $T_c$, it will quickly go across the unity $1$ and become divergent. In comparison with the works based on a weak first-order quark-hadron phase transition\cite{Scavenius:2000bb,Bessa:2008nw}, we find the $\Delta F_b/T$ as a function of temperature shows similar behavior when the temperature is nearby the critical temperature $T_c$, however, when the temperature is apart from $T_c$, our results present a non-monotonic behavior with the decrease of the temperature, whereas for a weak first-order phase transition with the spinodal instability,the $\Delta F_b/T$ will drop monotonically to zero very soon as $T\rightarrow T_{sp}$. This is another apparent feature between the zero-temperature effective potential with and without a barrier. In the end $\Delta F_b/T \simeq1$ corresponds to the moment when the system is likely to stay in the metastable vacuum for a relatively long time.

\begin{acknowledgments}
We thank Jinshuang Jin for valuable comments and discussions. This work is supported in part by National Natural Science Foundation of China (NSFC) under No.11675048..
\end{acknowledgments}


\begin{thebibliography}{199}

%\cite{Yagi:2005yb}
\bibitem{Yagi:2005yb}
  K.~Yagi, T.~Hatsuda and Y.~Miake,
  ``\textit{Quark-gluon plasma: From big bang to little bang},''
  Camb.\ Monogr.\ Part.\ Phys.\ Nucl.\ Phys.\ Cosmol.\  {\bf 23}, 1 (2005).  %%CITATION = CMPCE,23,1;%%

%\cite{Fukushima:2010bq}
\bibitem{Fukushima:2010bq}
  K.~Fukushima and T.~Hatsuda,
  %``The phase diagram of dense QCD,''
  Rept.\ Prog.\ Phys.\  {\bf 74}, 014001 (2011).  %%CITATION = ARXIV:1005.4814;%%

%\cite{Braun-Munzinger:2015hba}
\bibitem{Braun-Munzinger:2015hba}
  P.~Braun-Munzinger, V.~Koch, T.~Schäfer and J.~Stachel,
  %``Properties of hot and dense matter from relativistic heavy ion collisions,''
  Phys.\ Rept.\  {\bf 621}, 76 (2016).

   %\cite{Nambu:1961tp}
\bibitem{Nambu:1961tp}
  Y.~Nambu and G.~Jona-Lasinio,
  %``Dynamical Model of Elementary Particles Based on an Analogy with Superconductivity. 1.,''
  Phys.\ Rev.\  {\bf 122}, 345 (1961);
    %\cite{Nambu:1961fr}
%\bibitem{Nambu:1961fr}
  Y.~Nambu and G.~Jona-Lasinio,
  %``Dynamical Model Of Elementary Particles Based On An Analogy With Superconductivity. Ii,''
  Phys.\ Rev.\  {\bf 124}, 246 (1961).

  %\cite{Vogl:1991qt}
\bibitem{Vogl:1991qt}
  U.~Vogl and W.~Weise,
  %``The Nambu and Jona Lasinio model: Its implications for hadrons and nuclei,''
  Prog.\ Part.\ Nucl.\ Phys.\  {\bf 27}, 195 (1991);  %%CITATION = PPNPD,27,195;%%
%\cite{Klevansky:1992qe}
%\bibitem{Klevansky:1992qe}
  S.~P.~Klevansky,
  %``The Nambu-Jona-Lasinio model of quantum chromodynamics,''
  Rev.\ Mod.\ Phys.\  {\bf 64}, 649 (1992);  %%CITATION = RMPHA,64,649;%%
%\cite{Hatsuda:1994pi}
%\cite{Hatsuda:1994pi}
%\bibitem{Hatsuda:1994pi}
  T.~Hatsuda and T.~Kunihiro,
  %``QCD phenomenology based on a chiral effective Lagrangian,''
  Phys.\ Rept.\  {\bf 247}, 221 (1994);  %%CITATION = HEP-PH/9401310;%%
%\cite{Buballa:2003qv}
%\bibitem{Buballa:2003qv}
  M.~Buballa,
  %``NJL model analysis of quark matter at large density,''
  Phys.\ Rept.\  {\bf 407}, 205 (2005).  %%CITATION = HEP-PH/0402234;%%

 %\cite{GellMann:1960np}
\bibitem{GellMann:1960np}
  M.~Gell-Mann and MLevy,
  %``The axial vector current in beta decay,''
  Nuovo Cim.\  {\bf 16}, 705 (1960).  %%CITATION = NUCIA,16,705;%%

 %\cite{Costa:2010zw}
\bibitem{Costa:2010zw}
  P.~Costa, M.~C.~Ruivo, C.~A.~de Sousa and H.~Hansen,
  %``Phase diagram and critical properties within an effective model of QCD: the Nambu-Jona-Lasinio model coupled to the Polyakov loop,''
  Symmetry {\bf 2}, 1338 (2010), and references therein.

    %\cite{Schaefer:2007pw}
\bibitem{Schaefer:2007pw}
  B.~J.~Schaefer, J.~M.~Pawlowski and J.~Wambach,
  %``The Phase Structure of the Polyakov--Quark-Meson Model,''
  Phys.\ Rev.\  D {\bf 76}, 074023 (2007).

%\cite{Luo:2017faz}
\bibitem{Luo:2017faz}
  X.~Luo and N.~Xu,
  %``Search for the QCD Critical Point with Fluctuations of Conserved Quantities in Relativistic Heavy-Ion Collisions at RHIC : An Overview,''
  Nucl.\ Sci.\ Tech.\  {\bf 28}, 112 (2017).

%\cite{Aggarwal:2010cw}
\bibitem{Aggarwal:2010cw}
  M.~M.~Aggarwal {\it et al.} [STAR Collaboration],
  %``An Experimental Exploration of the QCD Phase Diagram: The Search for the Critical Point and the Onset of De-confinement,''
  arXiv:1007.2613 [nucl-ex].

%\cite{Abgrall:2014xwa}
\bibitem{Abgrall:2014xwa}
  N.~Abgrall {\it et al.} [NA61 Collaboration],
  %``NA61/SHINE facility at the CERN SPS: beams and detector system,''
  JINST {\bf 9}, P06005 (2014).

  %\cite{Herbst:2013ail}
\bibitem{Herbst:2013ail}
  T.~K.~Herbst, J.~M.~Pawlowski and B.~J.~Schaefer,
  %``Phase structure and thermodynamics of QCD,''
  Phys.\ Rev.\ D {\bf 88}, no. 1, 014007 (2013).

%\cite{Glendenning:2000}
\bibitem{Glendenning:2000}
N. Glendenning, ``\textit{Compact Stars. Nuclear Physics,Particle Physics, and General Relativity}'', Springer-Verlag, Berlin, (2000).

%\cite{Ferreira:2020evu}
\bibitem{Ferreira:2020evu}
  M.~Ferreira, R.~Câmara Pereira and C.~Providência,
  %``Neutron stars with large quark cores,''
  arXiv:2005.10543 [nucl-th].

%\cite{Xia:2020brt}
\bibitem{Xia:2020brt}
  C.~J.~Xia, T.~Maruyama, N.~Yasutake, T.~Tatsumi, H.~Shen and H.~Togashi,
  %``Systematic study on the quark-hadron mixed phase in compact stars,''
  arXiv:2005.02273 [hep-ph].

%\cite{Cao:2018tzm}
\bibitem{Cao:2018tzm}
G.~Cao and S.~Lin,
%``Gravitational Wave from Phase Transition inside Neutron Stars,''
[arXiv:1810.00528 [nucl-th]].
%1 citations counted in INSPIRE as of 10 Sep 2020

%\cite{Paschalidis:2017qmb}
\bibitem{Paschalidis:2017qmb}
  V.~Paschalidis, K.~Yagi, D.~Alvarez-Castillo, D.~B.~Blaschke and A.~Sedrakian,
  %``Implications from GW170817 and I-Love-Q relations for relativistic hybrid stars,''
  Phys.\ Rev.\ D {\bf 97}, no. 8, 084038 (2018)
%  doi:10.1103/PhysRevD.97.084038
  [arXiv:1712.00451 [astro-ph.HE]].

%\cite{Trodden:1998ym}
\bibitem{Trodden:1998ym}
M.~Trodden,
%``Electroweak baryogenesis,''
Rev. Mod. Phys. \textbf{71}, 1463-1500 (1999)
%doi:10.1103/RevModPhys.71.1463
[arXiv:hep-ph/9803479 [hep-ph]].

%\cite{Huang:2017kzu}
\bibitem{Huang:2017kzu}
F.~P.~Huang and C.~S.~Li,
%``Probing the baryogenesis and dark matter relaxed in phase transition by gravitational waves and colliders,''
Phys. Rev. D \textbf{96}, no.9, 095028 (2017)
%doi:10.1103/PhysRevD.96.095028
[arXiv:1709.09691 [hep-ph]].

%\cite{Ellis:2020awk}
\bibitem{Ellis:2020awk}
  J.~Ellis, M.~Lewicki and J.~M.~No,
  %``Gravitational waves from first-order cosmological phase transitions: lifetime of the sound wave source,''
  arXiv:2003.07360 [hep-ph].

%\cite{Wang:2020jrd}
\bibitem{Wang:2020jrd}
  X.~Wang, F.~P.~Huang and X.~Zhang,
  %``Phase transition dynamics and gravitational wave spectra of strong first-order phase transition in supercooled universe,''
  JCAP {\bf 2005}, 045 (2020)
 % doi:10.1088/1475-7516/2020/05/045
  [arXiv:2003.08892 [hep-ph]].

%\cite{Caprini:2019egz}
\bibitem{Caprini:2019egz}
C.~Caprini, M.~Chala, G.~C.~Dorsch, M.~Hindmarsh, S.~J.~Huber, T.~Konstandin, J.~Kozaczuk, G.~Nardini, J.~M.~No, K.~Rummukainen, P.~Schwaller, G.~Servant, A.~Tranberg and D.~J.~Weir,
%``Detecting gravitational waves from cosmological phase transitions with LISA: an update,''
JCAP \textbf{03}, 024 (2020)
%doi:10.1088/1475-7516/2020/03/024
[arXiv:1910.13125 [astro-ph.CO]].

%\cite{Coleman:1977py}
\bibitem{Coleman:1977py}
S.~R.~Coleman,
%``The Fate of the False Vacuum. 1. Semiclassical Theory,''
Phys. Rev. D \textbf{15}, 2929-2936 (1977)
doi:10.1103/PhysRevD.16.1248.

%\cite{Callan:1977pt}
\bibitem{Callan:1977pt}
C.~G.~Callan, Jr. and S.~R.~Coleman,
%``The Fate of the False Vacuum. 2. First Quantum Corrections,''
Phys. Rev. D \textbf{16}, 1762-1768 (1977)
doi:10.1103/PhysRevD.16.1762.

%\cite{Coleman:1988}
\bibitem{Coleman:1988}
S. Coleman, ``\textit{Aspects of Symmetry},'' Cambridge University Press, Cambridge, England, 1988. P416.

%\cite{Linde:1980tt}
\bibitem{Linde:1980tt}
  A.~D.~Linde,
  %``Fate of the False Vacuum at Finite Temperature: Theory and Applications,''
  Phys.\ Lett.\  {\bf 100B}, 37 (1981).
%  doi:10.1016/0370-2693(81)90281-1

%\cite{Linde:1981zj}
\bibitem{Linde:1981zj}
  A.~D.~Linde,
  %``Decay of the False Vacuum at Finite Temperature,''
  Nucl.\ Phys.\ B {\bf 216}, 421 (1983)
  Erratum: [Nucl.\ Phys.\ B {\bf 223}, 544 (1983)].
%  doi:10.1016/0550-3213(83)90293-6, 10.1016/0550-3213(83)90072-X

%\cite{Goldenfeld:1992qy}
\bibitem{Goldenfeld:1992qy}
  N.~Goldenfeld,
  ``\textit{Lectures on phase transitions and the renormalization group},''
 Addison-Wesley (1992)(Frontiers in physics, 85).

%\cite{Enqvist:1991xw}
\bibitem{Enqvist:1991xw}
  K.~Enqvist, J.~Ignatius, K.~Kajantie and K.~Rummukainen,
  %``Nucleation and bubble growth in a first order cosmological electroweak phase transition,''
  Phys.\ Rev.\ D {\bf 45}, 3415 (1992).
  doi:10.1103/PhysRevD.45.3415

%\cite{Scavenius:2000bb}
\bibitem{Scavenius:2000bb}
  O.~Scavenius, A.~Dumitru, E.~S.~Fraga, J.~T.~Lenaghan and A.~D.~Jackson,
  %``First order chiral phase transition in high-energy collisions: Can nucleation prevent spinodal decomposition?,''
  Phys.\ Rev.\ D {\bf 63}, 116003 (2001)
%  doi:10.1103/PhysRevD.63.116003
  [hep-ph/0009171].

%\cite{Bessa:2008nw}
\bibitem{Bessa:2008nw}
  A.~Bessa, E.~S.~Fraga and B.~W.~Mintz,
  %``Phase conversion in a weakly first-order quark-hadron transition,''
  Phys.\ Rev.\ D {\bf 79}, 034012 (2009)
 % doi:10.1103/PhysRevD.79.034012
  [arXiv:0811.4385 [hep-ph]].

%\cite{Palhares:2010be}
\bibitem{Palhares:2010be}
  L.~F.~Palhares and E.~S.~Fraga,
  %``Droplets in the cold and dense linear sigma model with quarks,''
  Phys.\ Rev.\ D {\bf 82}, 125018 (2010)
 % doi:10.1103/PhysRevD.82.125018
  [arXiv:1006.2357 [hep-ph]].

%\cite{Kroff:2014qxa}
\bibitem{Kroff:2014qxa}
  D.~Kroff and E.~S.~Fraga,
  %``Nucleating quark droplets in the core of magnetars,''
  Phys.\ Rev.\ D {\bf 91}, no. 2, 025017 (2015)
%  doi:10.1103/PhysRevD.91.025017
  [arXiv:1409.7026 [hep-ph]].

%\cite{Friedberg:1976eg}
\bibitem{Friedberg:1976eg}
  R.~Friedberg and T.~D.~Lee,
  %``Fermion Field Nontopological Solitons. 1,''
  Phys.\ Rev.\ D {\bf 15}, 1694 (1977); R.~Friedberg and T.~D.~Lee,
  %``Fermion Field Nontopological Solitons. 2. Models For Hadrons,''
  Phys.\ Rev.\ D {\bf 16}, 1096 (1977); R.~Friedberg and T.~D.~Lee,
  %``QCD And The Soliton Model Of Hadrons,''
  Phys.\ Rev.\ D {\bf 18}, 2623 (1978).

%\cite{Reinhardt:1985nq}
\bibitem{Reinhardt:1985nq}
  H.~Reinhardt, B.~V.~Dang and H.~Schulz,
  %``Deconfinement Phase Transition Of Hot And Dense Nuclear Matter In The Nontopological Soliton Bag Model,''
  Phys.\ Lett.\  {\bf 159B}, 161 (1985).
 % doi:10.1016/0370-2693(85)90878-0

%\cite{Li:1987wb}
\bibitem{Li:1987wb}
  M.~Li, M.~C.~Birse and L.~Wilets,
  %``Phase Transition In The Soliton Bag Model,''
  J.\ Phys.\ G {\bf 13} (1987) 1.

  %\cite{Gao:1992zd}
\bibitem{Gao:1992zd}
  S.~Gao, E.~K.~Wang and J.~R.~Li,
  %``Bag constant and deconfinement phase transition in a nontopological soliton
  %model,''
  Phys.\ Rev.\ D {\bf 46}, 3211 (1992).

%\cite{Mao:2007gm}
\bibitem{Mao:2007gm}
  H.~Mao, M.~Yao and W.~Q.~Zhao,
  %``The Friedberg-Lee model at finite temperature and density,''
  Phys.\ Rev.\ C {\bf 77}, 065205 (2008)
%  doi:10.1103/PhysRevC.77.065205
  [arXiv:0711.4643 [hep-ph]].

%\cite{Shu:2010xj}
\bibitem{Shu:2010xj}
  S.~Shu and J.~R.~Li,
  %``Describing the strongly interacting quark-gluon plasma through the Friedberg-Lee model,''
  Phys.\ Rev.\ C {\bf 82}, 045203 (2010)
 % doi:10.1103/PhysRevC.82.045203
  [arXiv:1003.2246 [hep-ph]].

  %\cite{Mao:2013qu}
\bibitem{Mao:2013qu}
H.~Mao, T.~Wei and J.~Jin,
%``Chiral soliton model at finite temperature and density,''
Phys. Rev. C \textbf{88}, 035201 (2013)
doi:10.1103/PhysRevC.88.035201
[arXiv:1301.6227 [hep-ph]].


%\cite{Jin:2015goa}
\bibitem{Jin:2015goa}
  J.~Jin and H.~Mao,
  %``Nontopological Soliton in the Polyakov Quark Meson Model,''
  Phys.\ Rev.\ C {\bf 93}, no. 1, 015202 (2016)
%  doi:10.1103/PhysRevC.93.015202
  [arXiv:1508.03920 [hep-ph]].
  %%CITATION = doi:10.1103/PhysRevC.93.015202;%%

%%%%%==============================================


%\cite{Goldflam:1981tg}
\bibitem{Goldflam:1981tg}
  R.~Goldflam and L.~Wilets,
  %``The Soliton Bag Model,''
  Phys.\ Rev.\  D {\bf 25}, 1951 (1982).

%\cite{Dolan:1973qd}
\bibitem{Dolan:1973qd}
  L.~Dolan and R.~Jackiw,
  %``Symmetry Behavior At Finite Temperature,''
  Phys.\ Rev.\ D {\bf 9}, 3320 (1974).

 %\cite{Weinberg:2012pjx}
\bibitem{Weinberg:2012pjx}
  E.~J.~Weinberg,
  ``\textit{Classical solutions in quantum field theory: Solitons and Instantons in High Energy Physics},''
 Cambridge Monographs on Mathematical Physics, Cambridge, (2012).
 % doi:10.1017/CBO9781139017787

%\cite{Gleiser:1993hf}
\bibitem{Gleiser:1993hf}
  M.~Gleiser, G.~C.~Marques and R.~O.~Ramos,
  %``On the evaluation of thermal corrections to false vacuum decay rates,''
  Phys.\ Rev.\ D {\bf 48}, 1571 (1993)
% doi:10.1103/PhysRevD.48.1571
  [hep-ph/9304234].

%\cite{Mintz:2012mz}
\bibitem{Mintz:2012mz}
  B.~W.~Mintz, R.~Stiele, R.~O.~Ramos and J.~Schaffner-Bielich,
  %``Phase diagram and surface tension in the three-flavor Polyakov-quark-meson model,''
  Phys.\ Rev.\ D {\bf 87}, no. 3, 036004 (2013)
% doi:10.1103/PhysRevD.87.036004
  [arXiv:1212.1184 [hep-ph]].

%\cite{Fraga:2018cvr}
\bibitem{Fraga:2018cvr}
  E.~S.~Fraga, M.~Hippert and A.~Schmitt,
  %``Surface tension of dense matter at the chiral phase transition,''
  Phys.\ Rev.\ D {\bf 99}, no. 1, 014046 (2019)
%  doi:10.1103/PhysRevD.99.014046
  [arXiv:1810.13226 [hep-ph]].

%\cite{Mao:2019aps}
\bibitem{Mao:2019aps}
Shen Wan-Ping, You Shi-Jia, Mao Hong,
%`` Phase structure and surface tension in quark meson model,''
Acta Physica Sinica, 2019, {\bf 68}(18): 181101.
%doi: 10.7498/aps.68.20190798

%\cite{Goldflam:1981tg}
\bibitem{Goldflam:1981tg}
  R.~Goldflam and L.~Wilets,
  %``The Soliton Bag Model,''
  Phys.\ Rev.\ D {\bf 25}, 1951 (1982).
 % doi:10.1103/PhysRevD.25.1951
  %%CITATION = doi:10.1103/PhysRevD.25.1951;%%

%\cite{Oertel:2008wr}
\bibitem{Oertel:2008wr}
  M.~Oertel and M.~Urban,
  %``Surface effects in color superconducting strangelets and strange stars,''
  Phys.\ Rev.\ D {\bf 77}, 074015 (2008)
%  doi:10.1103/PhysRevD.77.074015
  [arXiv:0801.2313 [nucl-th]].
  %%CITATION = doi:10.1103/PhysRevD.77.074015;%%

%\cite{Pinto:2012aq}
\bibitem{Pinto:2012aq}
  M.~B.~Pinto, V.~Koch and J.~Randrup,
  %``The Surface Tension of Quark Matter in a Geometrical Approach,''
  Phys.\ Rev.\ C {\bf 86}, 025203 (2012)
 % doi:10.1103/PhysRevC.86.025203
  [arXiv:1207.5186 [hep-ph]].
  %%CITATION = doi:10.1103/PhysRevC.86.025203;%%

%\cite{Garcia:2013eaa}
\bibitem{Garcia:2013eaa}
  A.~F.~Garcia and M.~B.~Pinto,
  %``Surface tension of magnetized quark matter,''
  Phys.\ Rev.\ C {\bf 88}, no. 2, 025207 (2013)
%  doi:10.1103/PhysRevC.88.025207
  [arXiv:1306.3090 [hep-ph]].
  %%CITATION = doi:10.1103/PhysRevC.88.025207;%%

%\cite{Ke:2013wga}
\bibitem{Ke:2013wga}
  W.~y.~Ke and Y.~x.~Liu,
  %``Interface tension and interface entropy in the 2+1 flavor Nambu-Jona-Lasinio model,''
  Phys.\ Rev.\ D {\bf 89}, no. 7, 074041 (2014)
%  doi:10.1103/PhysRevD.89.074041
  [arXiv:1312.2295 [hep-ph]].
  %%CITATION = doi:10.1103/PhysRevD.89.074041;%%

\end{thebibliography}
\end{document}